\documentclass[aps,preprint,preprintnumber,nofootinbib,amsmath,amssymb,ascmac,bm,12pt]{revtex4}
\usepackage{bm}
\usepackage{graphicx}
\usepackage{dcolumn}
\usepackage{bm}

\usepackage{color}

\newcommand{\bn}{\bar{n}}

\newcommand{\bvarphi}{\bar{\varphi}}

\newcommand{\br}{\bar{r}}
\newcommand{\btheta}{\bar{\theta}}

\newcommand{\lx}{\ell_1}
\newcommand{\ly}{\ell_2}
\newcommand{\lz}{\ell_3}

\newcommand{\omegac}{\omega_{\rm c}}

\begin{document}
\baselineskip5.5mm
\title{
Orbital angular momentum of scalar field \\
generated by gravitational scatterings
}
\bigskip
\author{
${}^{1}$Ryusuke Nishikawa,
${}^{1}$Ken-ichi Nakao \footnote{E-mail:knakao@sci.osaka-cu.ac.jp},
${}^{1}$Atsuki Masuda,
${}^{2}$Yasusada Nambu \footnote{E-mail:nambu@gravity.phys.nagoya-u.ac.jp},\\
and
${}^{1}$Hideki Ishihara \footnote{E-mail:ishihara@sci.osaka-cu.ac.jp},
\bigskip
}
\affiliation{
${}^{1}$Department of Mathematics and Physics,
Graduate School of Science, Osaka City University,
3-3-138 Sugimoto, Sumiyoshi, Osaka 558-8585, Japan
\\
${}^{2}$Department of Physics,
Graduate School of Science, Nagoya University, 
Chikusa, Nagoya 464-8602, Japan
\bigskip
}

\begin{abstract}
It has been expected that astronomical observations to detect 
the orbital angular momenta of electromagnetic waves may give us a new insight into astrophysics.
Previous works pointed out the possibility that 
a rotating black hole can produce orbital angular momenta 
of electromagnetic waves through gravitational scattering,  
and the spin parameter of the black hole can be measured by observing them. 
However, the mechanism how the orbital angular momentum of the electromagnetic wave is generated by 
the gravitational scattering has not been clarified 
sufficiently. In this paper, in order to understand it from a point of view of 
gravitational lensing effects, we consider an emitter which radiates a spherical wave 
of the real massless scalar field and study the deformation of  the scalar wave by 
the gravitational scattering due to a black hole by invoking the geometrical optics approximation. 
We show that the frame dragging caused by the rotating black hole is not a necessary condition 
for generating the orbital angular momentum of the scalar wave. However, its components 
parallel to the direction cosines of images appear only if the black hole is rotating.

\end{abstract}

\preprint{OCU-PHYS-451 }

\preprint{AP-GR-132}

\date{\today}
\maketitle

\section{introduction}\label{sec1}

The optical vortex is the electromagnetic wave which has a non-vanishing 
vorticity with respect to the spatial gradient of its phase function. 
It attracts attentions in optical physics especially after Allen and coworkers\cite{Allen:1992ci} 
have shown  that the Laguerre-Gaussian laser beam which is 
a typical example of optical vortex carries the orbital angular momentum 
clearly distinguishable from the spin angular momentum. 
In contrast to a particle, the Laguerre-Gaussian laser beam has an orbital 
angular momentum whose component in its propagation direction does not vanish.  
Such laser beams have been actively studied in many different fields\cite{Mair:2011ci,Wang:2012ci}. 

In general situations, the spin and orbital angular momenta of electromagnetic waves 
are not clearly distinguished. However, this fact does not necessarily deny the importance of 
the angular momenta of electromagnetic waves.  
Harwit~\cite{Harwit:2003ci} pointed out that astronomical observations 
to detect the orbital angular momenta of 
electromagnetic waves may give a new insight into astrophysics.
Previous works by Tamburini et al.~\cite{Tamburini:2011tk}, 
and Yang and Casals~\cite{Yang:2014kyr} showed that the orbital angular momentum of the 
electromagnetic wave can be produced by gravitational scattering of 
a rotating black hole, and the spectrum of a component of the orbital angular momentum  
can be a probe of the spin parameter of the black hole.
On the other hand, the mechanism how the orbital angular momentum  of the electromagnetic wave 
is generated by a black hole has not been well clarified yet. 
Hence, in this paper, we study the same subject as that in the previous 
studies \cite{Tamburini:2011tk} and \cite{Yang:2014kyr}, i.e., 
the generation of the orbital angular momentum  
through the gravitational scattering by a black hole. 

In this paper, for simplicity, we consider the classical real massless scalar field, 
since we are not interested in the spin angular momentum.  
We assume a situation in which an emitter radiates a monochromatic spherical wave, 
and then a part of it propagates through the vicinity of a black hole. 
We solve the equation of motion of the massless scalar field by invoking the 
geometrical optics approximation and then show that the generation 
of the orbital angular momentum  can be recognized as 
interferences caused by the gravitational lensing effects; The situation is very similar to  
the system studied by Masajada and Dubik \cite{Masajada:2001} in which 
the optical vortices can be generated by the interference of three plane waves.  
In contrast to the previous 
studies\cite{Tamburini:2011tk,Yang:2014kyr}, we investigate all components of 
the orbital angular momentum.   
Then, we will see that although the frame dragging 
due to the rotating black hole is not a necessary  
condition for the generation of the orbital angular momentum, 
its components parallel to the direction cosines of the images are produced only when  
the black hole is rotating. 

This paper is organized as follows. 
In \S~\ref{sec2}, we give a definition of the conserved quantities (the energy, the momentum and 
the orbital angular momentum)  of the real massless scalar field in the Minkowski spacetime. 
In \S~\ref{sec3}, we solve the equation of motion of the scalar field by 
the geometrical optics approximation and 
show that the scalar field gets a non-vanishing orbital angular momentum 
through the gravitational lensing effects due to a compact source of gravity. 
In \S~\ref{sec4}, we study the case in which the source of gravity is a black hole and  
show that through the measurements of components of the orbital angular momentum 
parallel to the direction cosines 
of images, we can know whether the black hole is rotating. 
\S~\ref{sec5} is devoted to the summary and the discussion.

In this paper, we use the geometrized units 
in which the speed of light $c$ and Newton's gravitational constant $G$ are one, 
but if necessary, those will be recovered. 
The Greek indices represent spacetime components, whereas the 
Latin indices represent spatial components.
We do not adopt the Einstein rule for the contraction of the Latin indices but of the Greek indices.   
The signature of the metric and sign convention of the Riemann tensor 
are the same as those of the textbook written by Wald\cite{Wald}.

\section{Conserved quantities in Minkowski spacetime}\label{sec2}

The action of the real massless scalar field $\Phi$ is given by 
\begin{equation}
S=-\int g^{\mu\nu} \left(\partial_\mu \Phi\right)\left(\partial_\nu\Phi\right) \sqrt{-g}~d^4x,
\end{equation}
where $\partial_\mu$ is the ordinary derivative with respect to $x^\mu$, $g^{\mu\nu}$ is the inverse of the 
metric tensor $g_{\mu\nu}$ of the spacetime, and $g$ is the determinant of $g_{\mu\nu}$. 
The minimum action principle leads to the equation of motion for $\Phi$ in the form 
\begin{equation}
\nabla^\mu\nabla_\mu\Phi=0,
\label{EOM}
\end{equation}
where $\nabla_\mu$ is the covariant derivative with respect to $g_{\mu\nu}$. 
The stress-energy tensor of $\Phi$ is given by
\begin{equation}
T_{\mu\nu}:=-\frac{1}{\sqrt{-g}}\frac{\delta S}{\delta g^{\mu\nu}}
=(\partial_\mu\Phi)\partial_\nu\Phi
-\frac{1}{2}g_{\mu\nu}g^{\alpha\beta}(\partial_\alpha\Phi)\partial_\beta\Phi 
\end{equation}
By using Eq.~(\ref{EOM}), we can see that $T^{\mu\nu}$ satisfies the local conservation law
\begin{equation}
\nabla_\mu T^{\mu\nu}=0. \label{T-cons}
\end{equation}

The infinitesimal coordinate transformation generated by the 
vector field $\xi^\mu$,   
\begin{equation}
\bar{x}^\mu=x^\mu+\varepsilon\xi^\mu, \label{c-trans}
\end{equation}
leads to the changes in the components of the metric tensor as
\begin{equation}
\bar{g}_{\mu\nu}(x^\alpha)=g_{\mu\nu}(x^\alpha)
- \varepsilon\left(\nabla_\mu\xi_\nu+\nabla_\nu\xi_\mu\right).
\label{m-trans}
\end{equation}
If $\xi^\mu$ satisfies 
\begin{equation}
\nabla_\mu\xi_\nu+\nabla_\nu\xi_\mu=0, \label{Killing-eq}
\end{equation}
then the transformation (\ref{c-trans}) 
does not change the components of the metric tensor. 
This fact implies that the vector field $\xi^\mu$ satisfying Eq.~(\ref{Killing-eq})  
is related to a symmetry of the spacetime, which is called the isometry. Equation (\ref{Killing-eq}) is 
called the Killing equation, and the solution of the Killing equation 
is called the Killing vector.  

If there is a Killing vector $\xi^\mu$ in the spacetime, 
Eqs.~(\ref{T-cons}) and (\ref{Killing-eq}) guarantee that 
the vector field $J^\mu$ defined as 
\begin{equation}
J^\mu:=T^\mu{}_\nu\xi^\nu 
\end{equation}
satisfies the conservation law
\begin{equation}
\partial_\mu \left(\sqrt{-g}J^\mu\right)=0. \label{c-law}
\end{equation}
If $J^\mu$ has a compact support, the following quantity $Q$ is a conserved quantity;
\begin{equation}
Q=\int_V J^0\sqrt{-g}d x^1 dx^2 dx^3, \label{c-quantity}
\end{equation}
where the integral is taken over the domain $V$ covering the support of $J^\mu$. 
From Eq.~(\ref{c-quantity}), we may regard $J^0\sqrt{-g}$ as the density of $Q$.  

In this paper, we consider the detection of the scalar field in the domain which is well described by 
the Minkowski geometry. Hence, we introduce conserved quantities associated with the Killing vectors 
in the Minkowski spacetime whose metric tensor is given by 
$g_{\mu\nu}={\rm diag}[-1,1,1,1]$ in the Cartesian inertial coordinates. 
Hereafter, we assume that the detector occupies a domain of 
$-\lx < x^1 <+\lx$, $-\ly < x^2 <+ \ly$ and $-\lz < x^3 <+\lz$. The conserved quantities 
in the detector are obtained by the integral over this domain.

\subsection{Energy}

The time coordinate basis is the timelike Killing vector; 
\begin{equation}
\xi_{\rm [0]}^\mu=\left(1,0,0,0\right). 
\end{equation}
Since the Killing vector $\xi_{\rm [0]}^\mu$ generates the time translation, 
the conserved quantity associated with it is the energy. 
The energy density $\rho_{\rm E}$ is defined as 
\begin{equation}
\rho_{\rm E}:=-T^0{}_\mu \xi_{\rm [0]}^\mu\sqrt{-g}
=\frac{1}{2}\left[\left(\partial_t\Phi\right)^2+\sum_{i=1}^3\left(\partial_i\Phi\right)^2\right]. 
\label{E-density}
\end{equation}
The energy $E$ in the detector is then given by
\begin{equation}
E=\int_{-\lz}^{+\lz}\int_{-\ly}^{+\ly}\int_{-\lx}^{+\lx} \rho_{\rm E}dx^1dx^2dx^3. \label{E-d}
\end{equation}

\subsection{Momentum}

The spatial coordinate basis vectors are the spatially translational Killing vectors; 
\begin{equation}
\xi_{\rm [1]}^\mu=\left(0,1,0,0\right),~~~~
\xi_{\rm [2]}^\mu=\left(0,0,1,0\right),~~~~
\xi_{\rm [3]}^\mu=\left(0,0,0,1\right). 
\end{equation}
The conserved quantity associated with these Killing vectors are the 
components of the momentum. The densities of the components of the momentum are defined as
\begin{equation}
\left(p^1,~p^2,~p^3\right)
:=\left(T^0{}_\mu\xi_{\rm [1]}^\mu,~ T^0{}_\mu\xi_{\rm [2]}^\mu ,~ T^0{}_\mu\xi_{\rm [3]}^\mu \right) 
=-\left(\partial_0\phi\right)\Bigl(\partial_1\phi,~\partial_2\phi,~\partial_3\phi \Bigr). \label{p-density}
\end{equation}
Then the momentum in the detector is given by
\begin{equation}
P^i=\int_{-\lz}^{+\lz}\int_{-\ly}^{+\ly}\int_{-\lx}^{+\lx} p^i ~dx^1dx^2dx^3, \label{P-d}
\end{equation}
where $i=1,2,3$. 

\subsection{Orbital angular momentum (OAM)}

The Killing vectors which generate the isometry of $SO(3)$ are
\begin{equation}
\xi_{\rm [2,3]}^\mu=\left(0,0,-x^3,x^2\right),~~~~
\xi_{\rm [3,1]}^\mu=\left(0,x^3,0,-x^1\right),~~~~
\xi_{\rm [1,2]}^\mu=\left(0,-x^2,x^1,0\right).
\end{equation}
The conserved quantity associated with these Killing vectors are the 
components of the orbital angular momentum (OAM). 
The densities of the components of the OAM are defined as
\begin{align}
\left(l^1,~l^2,~l^3\right)
&:=\left(T^0{}_\mu\xi_{\rm [2,3]}^\mu,~ T^0{}_\mu\xi_{\rm [3,1]}^\mu ,~ T^0{}_\mu\xi_{\rm [1,2]}^\mu \right) \cr
&=\left(x^2 p^3-x^3p^2,~x^3 p^1-x^1p^3,~x^1p^2-x^2p^1\right). \label{l-density}
\end{align}
Then, the OAM in the detector is given by
\begin{equation}
L^i=\int_{-\lz}^{+\lz}\int_{-\ly}^{+\ly}\int_{-\lx}^{+\lx} l^i ~dx^1dx^2dx^3. \label{L-d}
\end{equation}

\section{Generation of orbital angular momentum}\label{sec3}

We assume that a real scalar wave propagates in the stationary spacetime with {\it a 
compact source of gravity} ${\cal S}$; a spherical real scalar wave   
is emitted at a place very far from $\cal S$, 
a part of the spherical scalar wave propagates in the vicinity of $\cal S$ 
and then is detected at another place very far from $\cal S$. 
Hereafter, we call the domain 
where the scalar field is emitted {\it the emission domain} $\cal E$, whereas we call the domain 
where the scalar field is detected {\it the detection domain} ${\cal D}$. 

We assume that the emission domain $\cal E$ is well described by the Minkowski geometry, 
and set a spherical polar coordinate system $(t,\br,\btheta,\bvarphi)$, whose origin is the emitter. 
Then, as mentioned, we assume that a spherical scalar wave given below is emitted 
in the emission domain $\cal E$; We introduce the quantity defined as
\begin{equation}
\Psi|_{\cal E}=\frac{{\cal A} e^{-i\omegac (t-\br)}}{\br}, \label{initial-Psi}
\end{equation}
where ${\cal A}$ and $\omegac$ are real positive constants, 
and the scalar field $\Phi$ is the real part of $\Psi$. 
The scalar field $\Phi$ evolves in accordance with Eq.~(\ref{EOM}) and hence its functional form 
will be different from Eq.(\ref{initial-Psi}) in the detection domain $\cal D$. However, it is, in general, 
difficult to solve Eq.~(\ref{EOM}) in a curved spacetime, we invoke the geometrical optics 
approximation shown below. 

The geometrical optics approximation is available to 
the massless scalar field $\Phi$ governed by Eq.~(\ref{EOM}), 
if the following conditions hold (see, for example Ref.~\cite{SEF});
\begin{enumerate}
 \item The wavelength $\lambda$ is much shorter than the scale of 
the amplitude variation $\ell_{\rm A}$.
 \item The wavelength $\lambda$ is much shorter than 
 the spacetime curvature radius $\ell_{\rm C}$.
\end{enumerate}
These conditions imply that the non-negative  
parameter $\epsilon$ defined below is much less than unity; 
\begin{equation}
\epsilon=\left\{
\begin{array}{ll}
\displaystyle{\frac{\lambda}{\ell_{\rm A}}} &\quad {\rm for}~~\ell_{\rm A}< \ell_{\rm C}, \\
&\\
\displaystyle{\frac{\lambda}{\ell_{\rm C}}} &\quad {\rm for}~~\ell_{\rm C} \leq \ell_{\rm A}. \\
\end{array}\right.
\label{epsilon}
\end{equation}
Then, we may write the scalar field of the complex form $\Psi$ as 
\begin{eqnarray}
 \Psi=A(x^\mu)e^{i\frac{S(x^\mu)}{\epsilon}},
 \label{go-form}
\end{eqnarray}
where $A$ and $S$ are, respectively, the amplitude and the phase whose  
scales of variation are, by their definitions,  
\begin{eqnarray}
 \frac{\partial_\mu A}{A}=\mathcal{O}\left(\frac{1}{\ell_{\rm A}}\right)
 ~~{\rm and}~~
\frac{\partial_\mu S}{S}=\mathcal{O}\left(\frac{1}{\lambda}\right).
 \label{magnitude}
\end{eqnarray}
By substituting Eq.~\eqref{go-form} into the field equation~\eqref{EOM} and assuming the equation holds at each order with respect to $\epsilon$, the equation of the order $\epsilon^{-2}$ leads to the so-called eikonal equation:
\begin{eqnarray}
 g^{\mu\nu}\left(\partial_\mu S\right)\partial_\nu S=0.
 \label{eikonal}
\end{eqnarray}
By defining the 4-dimensional wave vector field $k_\mu$ as $k_\mu =\partial_\mu S$, 
the eikonal equation~\eqref{eikonal} leads to the null condition of the 4-dimensional wave vector:
\begin{eqnarray}
 k^\mu k_\mu =0,
 \label{null}
\end{eqnarray}
where as usual $k^\mu =g^{\mu\nu}k_\nu$.
By defining an affine parameter $\tau$ along an integral curve of the 4-dimensional wave vector as
\begin{eqnarray}
 \frac{d}{d\tau}=k^\mu \frac{\partial}{\partial x^\mu},
\end{eqnarray}
and by differentiating Eq.~\eqref{eikonal} with respect to $x^\nu$,  
we obtain the geodesic equations:
\begin{eqnarray}
 k^\nu \nabla_\nu k^\mu=0.
 \label{geodesic}
\end{eqnarray}
The equation of motion \eqref{EOM} of the next order with respect to $\epsilon$ 
leads to the so-called transport equation:
\begin{eqnarray}
 \frac{d}{d\tau}\ln A=-\frac{1}{2}\nabla_\mu k^\mu.
 \label{transport}
\end{eqnarray}
Equations~\eqref{null}, \eqref{geodesic} and \eqref{transport} are the basic equations to obtain 
the solution of Eq.~(\ref{EOM}) by the geometrical optics approximation. 
By solving the geodesic equations (\ref{geodesic}) with the null condition (\ref{null}), we obtain 
the congruence of the null geodesics or equivalently 
the null hypersurface with constant $S$. By solving Eq.~(\ref{transport}), we have 
$A$ along each null geodesic and hence  $\Psi$ through Eq.~(\ref{go-form}). 

As mentioned, we consider the situation in which the compact source of gravity $\cal S$ is 
located very far from the emitter in the emission domain $\cal E$; the distance between 
$\cal E$ and $\cal S$ is much larger than the curvature radius 
in the vicinity of $\cal S$. This fact implies that $\Psi$ well approximates to the plane waves 
in the vicinity of the compact source of gravity, and hence 
the geometrical optics approximation described above 
is available, if its wavelength is much shorter than the 
curvature length of the spacetime. 

The spatial size $\ell_{\cal D}$ of the detection domain ${\cal D}$ 
is assumed to be much larger than the wavelength $\lambda$ 
and much smaller than the scale of the amplitude variation $\ell_{\rm A}$  
and the spacetime curvature radius $\ell_{\rm C}$, 
that is, $\ell_{\cal D}$ satisfies
\begin{eqnarray}
 \lambda \ll \ell_{\cal D} \ll \frac{\lambda}{\epsilon}.
 \label{ell}
\end{eqnarray}
We assume that the detection domain $\cal D$ is located so distant from the source of gravity 
that the rest frame of the detector is almost a local inertial frame. 
If there is no source of gravity, $\Psi$ 
in the detection domain $\mathcal{D}$ is approximately written in the following form: 
\begin{eqnarray}
 \left.\Psi\right|_{\cal D} \simeq \Psi_{\rm plane}:=
 C e^{i\left(p_\mu y^\mu +\delta\right)},
 \label{pw1}
\end{eqnarray}
where $C$ is a real constant, and $y^\mu$, $p_\mu$ and $\delta$ denote 
the coordinates of the local inertial frame which covers the detection domain $\cal D$, 
the 4-dimensional wave vector, and the constant phase, respectively (see, Ref.~\cite{SEF}). 
The spatial coordinates $y^k$ ($k=1,2,3$) is assumed to be Cartesian. 
Hereafter, we call a wave well described by Eq.~\eqref{pw1} the   
{\it locally plane wave}. 

If the spacetime is stationary, $p_0$ is conserved, since $p^\mu$ satisfies the 
geodesic equations (\ref{geodesic}). We are interested in such cases, 
and hence hereafter we assume so and have $p_0=-\omegac$. 
There may be multiple null geodesics from the emitter to the detector 
in general curved spacetime due to the so-called gravitational lensing effects;  
The number of these null geodesics are equivalent to that of the images. 
Then, since the expression~\eqref{go-form} may be replaced by 
the superposition of locally plane waves (see, Ref.~\cite{SEF}), 
by denoting $y^\mu$ by $(t,\vec{y})$, $\Psi$ in the detection domain $\cal D$ is approximately given by 
\begin{eqnarray}
 \left.\Psi\right|_{\mathcal{D}} \simeq 
 \sum_{n=1}^N  C_{(n)}e^{-i\omegac\left(t-\vec{\gamma}_{(n)}\cdot \vec{y}\right)+i\delta_{(n)}},
 \label{multi-waves-c}
\end{eqnarray}
where $C_{(n)}$ is a real constant,  
each locally plane wave in the sum of the right hand side of Eq.~(\ref{multi-waves-c}) 
corresponds to one of the null geodesics from the emitter to 
the detector, the integer $N$ is the number of the null geodesics, and 
$\vec{\gamma}_{(n)}$ is the direction cosine of the $n$-th null geodesic. 
We will label each locally plane wave with a natural number so that $C_{(n)}\geq C_{(n+1)}$. 
The scalar field $\Phi$ is then given by
\begin{eqnarray}
 \left.\Phi\right|_{\cal D}=  \left.{\rm Re}\left[\Psi\right]\right|_{\cal D} \simeq  
 \sum_{n=1}^N  C_{(n)}  \cos \left[\omegac\left(t-\vec{\gamma}_{(n)}\cdot \vec{y}\right)-\delta_{(n)}\right],
 \label{multi-waves}
\end{eqnarray}
We should note that the phase difference $\delta_{(n)}-\delta_{(\bn)}$ with $n\neq \bn$ comes from the 
differences of the travel time $t_{(n)}-t_{(\bn)}$ and 
the number of caustics ${\cal N}_{(n)}-{\cal N}_{(\bn)}$ 
between the $n$-th and the $\bn$-th null geodesics; 
\begin{equation}
\delta_{(n)}-\delta_{(\bn)}=\omegac\left(t_{(n)}-t_{(\bn)}\right)
+\frac{\pi}{2}\left({\cal N}_{(n)}-{\cal N}_{(\bn)}\right).
\end{equation}
Hence, we should regard $\delta_{(n)}$ as a function of $\omegac$.

\subsection{Energy in the detector}

Substituting Eq.~(\ref{multi-waves}) into Eq.~(\ref{E-density}), we have 
\begin{align}
\rho_{\rm E}&=\frac{\omega_{\rm c}^2}{2}\sum_{n=1}^N\sum_{\bn=1}^N 
C_{(n)}C_{(\bn)}\left(1+\vec{\gamma}_{(n)}\cdot \vec{\gamma}_{(\bn)}\right) \cr
&\times
\sin\left[\omega_{\rm c}\left(t-\vec{\gamma}_{(n)}\cdot \vec{y}\right)-\delta_{(n)}\right]
\sin\left[\omega_{\rm c}\left(t-\vec{\gamma}_{(\bn)}\cdot\vec{y}\right)-\delta_{(\bn)}\right].
\end{align}
Through observations, we obtain the time average of $\rho_{\rm E}$ rather than itself; 
\begin{align}
\langle \rho_{\rm E}\rangle &\equiv \frac{\omega_{\rm c}}{2\pi}\int_T^{T+{2\pi\over\omega_{\rm c}}}
\rho_{\rm E} dt \cr
&=\frac{\omega_{\rm c}^2}{4}\sum_{n=1}^N\sum_{\bn=1}^N 
C_{(n)}C_{(\bn)}\left(1+\vec{\gamma}_{(n)}\cdot \vec{\gamma}_{(\bn)}\right) 
\cos\left[\omega_{\rm c}\left(\vec{\gamma}_{(n)}-\vec{\gamma}_{(\bn)}\right)\cdot\vec{y}
+\delta_{(n)}-\delta_{(\bn)}\right],
\end{align}
where $T$ is arbitrary time. 
Then, from Eq.~(\ref{E-d}), we have the time average of the energy in the detector in the form 
\begin{align}
\langle E \rangle 
&=\int_{-\lz}^{+\lz}\int_{-\ly}^{+\ly}\int_{-\lx}^{+\lx}\langle\rho_{\rm E} \rangle dy^1dy^2 dy^3 \cr
&=\frac{2}{\omega_{\rm c}}\sum_{n=1}^N\sum_{\bn=1}^N
\frac{C_{(n)}C_{(\bn)}\cos\left(\delta_{(n)}-\delta_{(\bn)}\right)}
{\varGamma^1_{(n,\bn)}\varGamma^2_{(n,\bn)}\varGamma^3_{(n,\bn)}} 
\left(1+\vec{\gamma}_{(n)}\cdot \vec{\gamma}_{(\bn)}\right)\cr
&\times \sin\left(\lx\omegac\varGamma^1_{(n,\bn)}\right)
\sin\left(\ly\omegac\varGamma^2_{(n,\bn)}\right)
\sin\left(\lz\omegac\varGamma^3_{(n,\bn)}\right),
\label{E-detect}
\end{align}
where
\begin{equation}
\varGamma_{(n,\bn)}^i\equiv \gamma_{(n)}^i -\gamma_{(\bn)}^i.
\end{equation}

\subsection{Momentum in the detector}

Substituting Eq.~(\ref{multi-waves}) into Eq.~(\ref{p-density}), we have 
\begin{align}
p^i&=\omega_{\rm c}^2\sum_{n=1}^N\sum_{\bn=1}^N 
C_{(n)}C_{(\bn)}\gamma^i_{(n)} \cr
&\times
\sin\left[\omega_{\rm c}\left(t-\vec{\gamma}_{(n)}\cdot \vec{y}\right)-\delta_{(n)}\right]
\sin\left[\omega_{\rm c}\left(t-\vec{\gamma}_{(\bn)}\cdot\vec{y}\right)-\delta_{(\bn)}\right].
\end{align}
Through observations, we obtain the time average of $p^i$; 
\begin{align}
\langle p^i \rangle &\equiv 
\frac{\omega_{\rm c}}{2\pi}\int_T^{T+{2\pi\over\omega_{\rm c}}}p^i ~dt \cr
&=\frac{\omega_{\rm c}^2}{2}\sum_{n=1}^N\sum_{\bn=1}^N 
C_{(n)}C_{(\bn)}\gamma^i_{(n)} 
\cos\left[\omega_{\rm c}\left(\vec{\gamma}_{(n)}-\vec{\gamma}_{(\bn)}\right)\cdot\vec{y}+\delta_{(n)}-\delta_{(\bn)}\right].
\end{align}
Then, from Eq.~(\ref{P-d}), we have the time average of the momentum in the detector in the form 
\begin{align}
\langle P^i \rangle 
&=\int_{-\lz}^{+\lz}\int_{-\ly}^{+\ly}\int_{-\lx}^{+\lx}\langle p^i \rangle dy^1dy^2dy^3 \cr
&= \frac{4}{\omega_{\rm c}}\sum_{n=1}^N\sum_{\bn=1}^N
\frac{C_{(n)}C_{(\bn)}\cos\left(\delta_{(n)}-\delta_{(\bn)}\right)}
{\varGamma^x_{(n,\bn)}\varGamma^y_{(n,\bn)}\varGamma^z_{(n,\bn)}} 
~\gamma^i_{(n)}\cr
&\times 
\sin\left(\lx\omegac\varGamma^1_{(n,\bn)}\right)
\sin\left(\ly\omegac\varGamma^2_{(n,\bn)}\right)
\sin\left(\lz\omegac\varGamma^3_{(n,\bn)}\right).
\label{P-detect}
\end{align}

\subsection{OAM in the detector}
 
Substituting Eq.~(\ref{multi-waves}) into Eq.~(\ref{l-density}), we have 
\begin{align}
l^i&=\omega_{\rm c}^2\sum_{n=1}^N\sum_{\bn=1}^N 
C_{(n)}C_{(\bn)}\sum_{j,k=1}^3\epsilon^{ijk}y^j\gamma^k_{(n)} \cr
&\times
\sin\left[\omega_{\rm c}\left(t-\vec{\gamma}_{(n)}\cdot \vec{y}\right)-\delta_{(n)}\right]
\sin\left[\omega_{\rm c}\left(t-\vec{\gamma}_{(\bn)}\cdot\vec{y}\right)-\delta_{(\bn)}\right],
\end{align}
where $\epsilon^{ijk}$ is the Levi-Civita symbol of $\epsilon^{123}=+1$. 
Through observations, we obtain the time average of $l^i$; 
\begin{align}
\langle l^i \rangle &\equiv \frac{\omega_{\rm c}}{2\pi}\int_T^{T+{2\pi\over\omega_{\rm c}}}l^i ~dt \cr
&=\frac{\omega_{\rm c}^2}{2}\sum_{n=1}^N\sum_{\bn=1}^N 
C_{(n)}C_{(\bn)}\sum_{j,k=1}^3\epsilon^{ijk}y^j \gamma^k_{(n)} 
\cos\left[\omega_{\rm c}\left(\vec{\gamma}_{(n)}-\vec{\gamma}_{(\bn)}\right)\cdot\vec{y}+\delta_{(n)}-\delta_{(\bn)}\right].
\end{align}
Then, from Eq.~(\ref{P-d}), we have the time average of the OAM in the detector in the form 
 \begin{align}
 \langle L^i\rangle &=\int_{-\lz}^{+\lz}\int_{-\ly}^{+\ly}\int_{-\lx}^{+\lx}\langle l^i \rangle dy^1dy^2dy^3 \cr
 & =\frac{4}{\omegac^2}\sum_{n=1}^N\sum_{\bn\neq n}^N 
 \frac{C_{(n)}C_{(\bn)}\sin\left(\delta_{(n)}-\delta_{(\bn)}\right)}
 {\varGamma_{(n,\bn)}^1 \varGamma_{(n,\bn)}^2\varGamma_{(n,\bn)}^3} 
 \sin\left( \ell_i\omegac\Gamma_{(n,\bn)}^i\right)\cr
 &\times
 \sum_{j,k=1}^3\epsilon^{ijk}
\gamma_{(n)}^k\sin\left(\ell_k\omegac\Gamma_{(n,\bn)}^k\right)
 \left[\omegac\ell_j \cos\left(\ell_j\omegac\varGamma_{(n,\bn)}^j\right)
 -\frac{\sin\left(\ell_j\omegac \varGamma_{(n,\bn)}^j\right)}{\varGamma_{(n,\bn)}^j}\right]. 
 \label{L-detect}
 \end{align}
Equation (\ref{L-detect}) implies that if there is only one image, $\langle L^i\rangle$ necessarily vanishes. 
The appearance of multiple images due to the gravitational lensing effect is a necessary condition 
of the generation of the OAM. 

\subsection{Average OAM of one quantum}

Although we consider the classical scalar field, it is useful for later discussion to 
introduce the number of quanta in the detector and the average OAM of one quantum. 
We may define the number of quanta in the detector as
$\langle E\rangle /\hbar \omega_{\rm c}$, where $\hbar$ is the Dirac constant. 
Hence, we have the average OAM that a quantum of the scalar field posseses as follows;
\begin{equation}
m^i\equiv \hbar \omegac\frac{ \langle L^i\rangle}{\langle E\rangle}. \label{OAM-q}
\end{equation}

\subsection{Some consequences}

Here we consider the non-trivial case 
that there are two or more images due to the gravitational lensing. 
Then, $\langle L^i \rangle$ will be a non-zero vector. 
However, in the case that the direction cosine of any image   
is written in terms of a linear combination of the direction cosines of any two images, 
$\langle L^i\rangle$ in the domain chosen in the manner given below 
is necessarily orthogonal to the direction cosines of all images. We show this fact below.  

 In the case we consider, direction cosines of all images span 
a 2-dimensional subspace of the 3-dimensional tangent space orthogonal to $dy^0$ 
at the detector.  Hereafter, we refer this subspace as $S$ 
and choose $\partial/\partial y^2$ and  $\partial/\partial y^3$ in $S$. 
Then, we consider $ \langle L^i\rangle$ in the domain, 
$-\ell_1< y^1 <+\ell_1$, $-\ell_2< y^2 <+\ell_2$ and $-\ell_3< y^3 <+\ell_3$.  
Since the direction cosines of images are observable, this experimental setting is, 
in principle, possible. 

For any element $w^i$ of $S$, we have
\begin{align}
\sum_{i=1}^3 w^i \langle l^i \rangle
&=\frac{\omega_{\rm c}^2}{2}\sum_{n=1}^N\sum_{\bn=1}^N 
C_{(n)}C_{(\bn)}\sum_{j,k=1}^3\epsilon^{i1k}w^i y^1\gamma^k_{(n)} \cr
&\times\cos\left[\omega_{\rm c}\left(\varGamma_{(n,\bn)}^2y^2+\varGamma_{(n,\bn)}^3y^3\right)
+\delta_{(n)}-\delta_{(\bn)}\right]. \label{proof-1}
\end{align}
By integrating both sides of Eq.~(\ref{proof-1}) with respect to $y^1$ over the domain $-\lx<y^1<+\lx$, 
we have
\begin{align}
\sum_{i=1}^3 w^i\int_{-\lx}^{+\lx}\langle l^i \rangle dy^1 
&=\frac{\omega_{\rm c}^2}{2}\sum_{n=1}^N\sum_{\bn=1}^N 
C_{(n)}C_{(\bn)}\sum_{j,k=1}^3\epsilon^{i1k}w^i\left(\int_{-\lx}^{+\lx}y^1dy^1\right)\gamma^k_{(n)} \cr
&\times\cos\left[\omega_{\rm c}\left(\varGamma_{(n,\bn)}^2y^2+\varGamma_{(n,\bn)}^3y^3\right)
+\delta_{(n)}-\delta_{(\bn)}\right] \cr
&=0. \label{proof-2}
\end{align}
Equation (\ref{proof-2}) implies 
\begin{equation}
\sum_{i=1}^3w^i \langle L^i \rangle =0. \label{Th-1}
\end{equation}
Furthermore, Eqs.~(\ref{P-detect}) and (\ref{Th-1}) implies 
\begin{equation}
\sum_{i=1}^3\langle P^i \rangle\langle L^i \rangle =0. \label{Th-2}
\end{equation}
As in the case of a particle, the OAM is orthogonal to the momentum. 
Masajada and Dubik have shown that the superposition of two plane waves  
does not generate optical vortex, but that of three plane waves does\cite{Masajada:2001}.  
This fact together with the present result implies that the non-vanishing orbital angular momentum 
does not necessarily imply the existence of a vortex.

\section{Scattering of scalar waves by a black hole}\label{sec4}

In this section, as mentioned, we consider the case that the compact source of gravity is a black hole. 

\subsection{The case of a rotating black hole}

The spacetime geometry of a stationary black hole is described by that of the Kerr
spacetime which is an exact solution of the vacuum Einstein equations. 
The infinitesimal world interval of the Kerr spacetime in the Boyer-Lindquist coordinates, 
$-\infty<r<\infty$, $0\leq \theta\leq\pi$, $0\leq \varphi <2\pi$, 
is given by
\begin{align}
 ds^2&=-\left(1-\frac{2Mr}{\Sigma}\right) dt^2
 -\frac{4aMr \sin^2\theta}{\Sigma}dtd\varphi
 +\Sigma d\theta^2+\frac{\Xi \sin^2\theta}{\Sigma}
 d\varphi^2
 +\frac{\Sigma}{\Delta}dr^2,
 \label{metric}
\end{align}
where 
\begin{align}
\Sigma&=r^2+a^2\cos^2\theta, \\
\Delta &=r^2-2Mr+a^2, \\
\Xi &=(r^2+a^2)^2-a^2\Delta\sin^2\theta,
\end{align}
and the two parameters, $M$ and $a$, denote the ADM mass and the specific angular momentum of this 
system, respectively. We assume that $M$ is non-negative. 
The parameter $a$ is often called the Kerr parameter. 
There is a spacetime singularity at $(r,\theta)=(0,\pi/2)$. 
In the case of $a^2\leq M^2$, the event horizon is located at $r=r_+:=M+\sqrt{M^2-a^2}$, 
whereas, in the case of $a^2>M^2$, there is no event horizon and the 
spacetime singularity is naked. In this subsection, we focus on the case of $a=0.99M$, 
the black hole. 

We assume an emitter on a world line specified by  
constant spatial coordinates $(r,\theta,\varphi)=(r_{\rm e},\theta_{\rm e},\varphi_{\rm e})$, which 
emits a massless scalar field described by the spherical scalar wave $\Psi$ given by  
Eq.~(\ref{initial-Psi}). We also assume a detector on a world line specified by 
constant spatial coordinates $(r,\theta,\varphi)=(r_{\rm d},\theta_{\rm d},\varphi_{\rm d})$.

In general, there are infinite number of null geodesics from the emitter to the detector 
in the Kerr spacetime, since there are null geodesics winding around a black hole boundlessly  
many times. Hence $\Psi$ in the vicinity of the detector 
is given by Eq.~\eqref{multi-waves} with $N=\infty$. 
Here we should note that 
the more the winding number of a null geodesic congruence around the black hole becomes, 
the more frequently the caustics occur on the null geodesic congruence. Due to the occurrence of 
caustics, $|C_{(n)}|\gg |C_{(n+1)}|$ holds for $n\geq2$. 
Thus, we consider only three locally plane waves $n\leq3$ that largely contribute to $\Psi$. 
Hereafter, we call these three null geodesics of $n=1,2,3$ 
the first, the second and the third ray, respectively. 

As mentioned, since the Kerr spacetime is stationary, we have 
\begin{eqnarray}
 k_t={\rm constant}=-\omega_{\rm c}.
 \label{frequency}
\end{eqnarray}

We put the emitter and the detector at the following 
points;
\begin{eqnarray}
 (\theta_{\rm e},\varphi_{\rm e},r_{\rm e})=(\pi/2,\pi,10^4M),~~{\rm and}~~
 (\theta_{\rm d},\varphi_{\rm d},r_{\rm d})\simeq (1.54978,1.98471\times10^{-2},10^4M).
 \label{locations}
\end{eqnarray}
Then, by numerically solving the geodesic equations, 
we obtain the world lines of three rays from the emitter to the detector.

\begin{figure}[htbp]
 \begin{center}
 \includegraphics[width=14cm,clip]{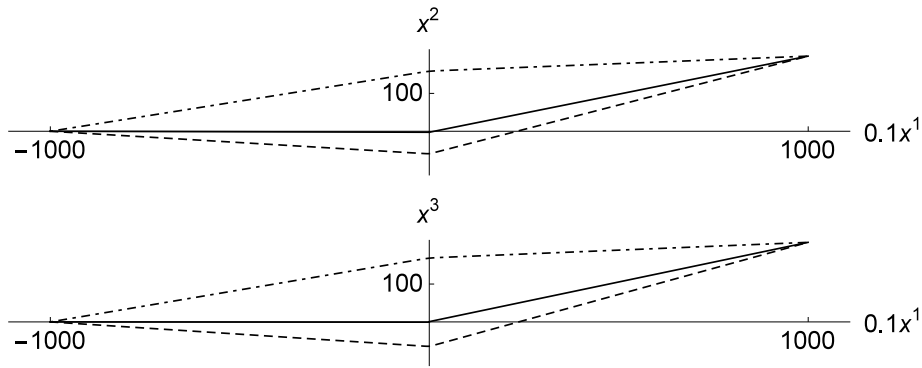}
 \end{center}
 \caption{
 A projection of the trajectories of first ray (dotdashed line), second ray (dashed line) 
 and third ray (solid line) on $(x'^1,x'^2)$ plane (upper panel) and $(x'^1,x'^3)$ plane (lower panel), 
 in the case of $a=0.99M$. 
 The horizontal and the vertical axes of each panels are $0.1x'^1$ and $x'^2,x'^3$.
 }
 \label{figrays}
\end{figure}
\begin{figure}[htbp]
 \begin{center}
 \includegraphics[width=12cm,clip]{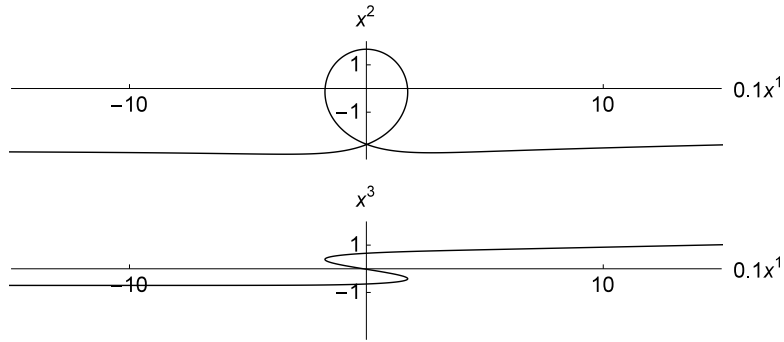}
 \end{center}
 \caption{
 A close-up of the vicinity of the black hole in Fig.~\ref{figrays} .
 }
 \label{figrayup}
\end{figure}
In Fig.~\ref{figrays}, we show a projection of the trajectories of the first ray (dot-dashed line), 
the second ray (dashed line) and the third ray (solid line) on $(x'^1,x'^2)$ 
plane and $(x'^1,x'^3)$ plane, 
where we have defined the Cartesian spatial coordinates $(x'^1,x'^2,x'^3)$ related to the Boyer-Lindquist spatial 
coordinates $(\theta,\varphi,r)$ through 
\begin{eqnarray}
 (x'^1,x'^2,x'^3)=(r\sin\theta\cos\varphi,r\sin\theta\sin\varphi,r\cos\theta).
 \label{xcoord}
\end{eqnarray}
In Fig.~\ref{figrayup}, we plot a close-up of the third ray in the vicinity of the black hole.
We can see that the third ray winds around the black hole one time.  

In order to know the interference of locally plane waves, we introduce the following orthonormal basis;
\begin{align}
e_{[t]}^\mu&=\frac{1}{\sqrt{-g^{tt}}}\left(1,0,0,0\right), \\
e_{[\theta]}^\mu&=\sqrt{g^{\theta\theta}}\left(0,1,0,0\right),\\
e_{[\varphi]}^\mu&=\frac{1}{\sqrt{g^{\varphi\varphi}}}\left(g^{t\varphi},0,g^{\varphi\varphi},0\right), \\
e_{[r]}^\mu&=\sqrt{g^{rr}}\left(0,0,0,1\right),
\end{align}
where $g^{\mu\nu}$ is the inverse of $g_{\mu\nu}$; 
\begin{align}
g^{tt}&=-\frac{\Xi}{\Sigma\Delta}, ~~~g^{t\varphi}=-\frac{2aMr}{\Sigma\Delta},~~~
g^{\theta\theta}=\frac{1}{\Sigma},~~~
g^{\varphi\varphi}=\frac{1}{\Delta\sin^2\theta}\left(1-\frac{2Mr}{\Sigma}\right),~~~
g^{rr}=\frac{\Delta}{\Sigma}. \nonumber
\end{align}
The unit vector $e_{[t]}^\mu$ is not normal to the spacelike hypersurface of constant $t$ but 
tangent to the time coordinate basis $\partial/\partial t$.   
This orthonormal basis may defines the local rest frame for the detector: 
$e_{[t]}^\mu$, $e_{[\theta]}^\mu$, $e_{[\varphi]}^\mu$ and $e_{[r]}^\mu$ correspond to 
$\partial/\partial y^0$, $\partial/\partial y^1$, $\partial/\partial y^2$ and $\partial/\partial y^3$, 
respectively. 

The numerical accuracy has been checked by investigating the null condition, i.e., whether 
$$
{\rm Err}:=\frac{2\left|-\left(k_{[t]}\right)^2+\left(k_{[\theta]}\right)^2
+\left(k_{[\varphi]}\right)^2+\left(k_{[r]}\right)^2\right|}
{\left(k_{[t]}\right)^2+\left(k_{[\theta]}\right)^2
+\left(k_{[\varphi]}\right)^2+\left(k_{[r]}\right)^2}
$$
is much less than unity, where 
$$
k_{[A]}=e_{[A]}^\mu k_\mu~~~~ {\rm with}~~A=\theta,\varphi,r.
$$ 
In our numerical calculations, the quantity Err at the detector 
is at most $10^{-6}$.

The direction cosine of the null geodesic in the local rest frame  
$(y^1,y^2,y^3)$ is given by  
\begin{align}
\gamma^i=\left|k_{[t]}\right|^{-1}\left(k_{[\theta]},~k_{[\varphi]},~k_{[r]}\right). 
\end{align}
The numerical results are 
\begin{align}
\gamma^i_{(1)}&=\left(1.68954\times10^{-2},~-1.59001\times10^{-2},~0.999731\right), \label{d-cosine1}\\
\gamma^i_{(2)}&=\left(-6.48922\times10^{-3},~5.96969\times10^{-3},~0.999962\right), \\
\gamma^i_{(3)}&=\left(7.02201\times10^{-5},~2.61991\times10^{-4},~0.999997\right). \label{d-cosine3}
\end{align}
The three direction cosines are linearly independent. This fact comes from 
the frame dragging due to the rotation of the black hole. 

In order to obtain the amplitudes of the locally plane waves, we solve the transport 
equation~\eqref{transport} along the three rays with an identical initial condition 
\begin{eqnarray}
 A^{-1}=0 \label{i-condition}
\end{eqnarray}
at the emitter (see e.g. Ref.~\cite{Holz:1997ic}, in details, 
for solving the transport equation). Note that the initial condition (\ref{i-condition}) 
is consistent with Eq.~(\ref{initial-Psi}).  
\begin{figure}[htbp]
 \begin{center}
 \includegraphics[width=8cm,clip]{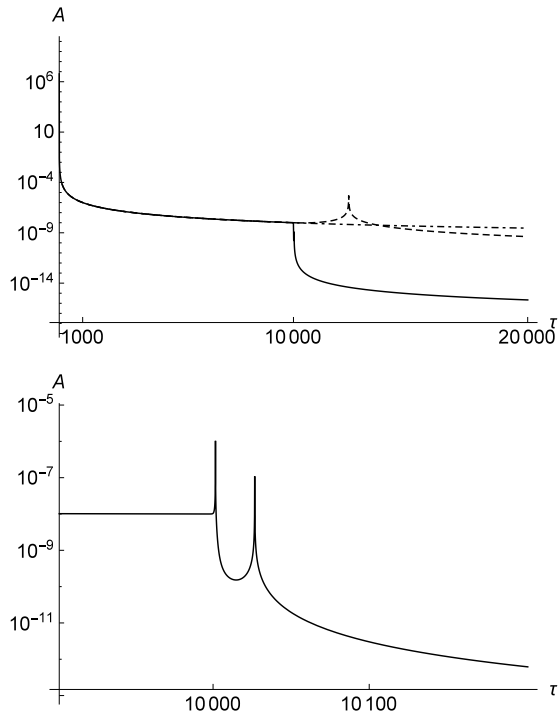}
 \end{center}
 \caption{
 The amplitudes along the trajectories of first ray (dot-dashed line), 
 second ray (dashed line) and third ray (solid line) as functions of $\tau$ from the emitter 
 to the detector. The lower panel is a close-up of the upper panel around the caustics of third ray.
 }
 \label{figAMP}
\end{figure}
In Fig.~\ref{figAMP}, we plot the amplitudes along the three 
rays as functions of $\tau$ from the source to the observer.
The lower panel is a close-up of the upper panel around the caustics of the third ray.
From Fig.~\ref{figAMP}, we can see that the second ray have one caustic and the third ray two.
From the numerical results, we have 
\begin{eqnarray}
 \left.\frac{A_{(2)}}{A_{(1)}}\right|_{t_{\rm d}}\simeq 1.42968\times 10^{-1},~~{\rm and}~~
 \left.\frac{A_{(3)}}{A_{(1)}}\right|_{t_{\rm d}}\simeq 7.39077\times 10^{-8},
\end{eqnarray}
where $A_{(1)}$, $A_{(2)}$ and $A_{(3)}$ denote the amplitudes along each ray. 
Hence, the ratios of the amplitudes $C_{(n)}$ in Eq.~\eqref{multi-waves} are given by 
\begin{eqnarray}
 \frac{C_{(2)}}{C_{(1)}}= 1.42968\times 10^{-1},~~~~{\rm and}~~~~
 \frac{C_{(3)}}{C_{(1)}}= 7.39077\times 10^{-8}.
 \label{amplitudes}
\end{eqnarray}

The phase $\delta_{(n)}$ in Eq.~\eqref{multi-waves} is 
determined from the travel time and the number of caustics (see, Ref~\cite{SEF});
\begin{eqnarray}
 \delta_{(n)}=\omega_{\rm c} \left(t_{(n)}(\tau_0)-t_{(n)}(0)\right)+\frac{\pi}{2}\mathcal{N}_{(n)},
\end{eqnarray}
where $t_{(n)}(\tau)$ represents the time coordinate along $n$-th ray, 
and $\mathcal{N}_{(n)}$ represents the number of caustics that $n$-th ray has. 
As mentioned, we have 
\begin{eqnarray}
 \mathcal{N}_{(1)}=0,~~\mathcal{N}_{(2)}=1,~~{\rm and}~~\mathcal{N}_{(3)}=2.
 \label{caustics}
\end{eqnarray}
Then, by using the numerical results and Eq.~\eqref{caustics}, we have the phase differences as follows; 
\begin{align}
 &\delta_{(2)}(\omega_{\rm c})-\delta_{(1)}(\omega_{\rm c})\simeq8.55591M\omega_{\rm c} +\frac{\pi}{2} 
 \label{constphase1}\\
 & \delta_{(3)}(\omega_{\rm c})-\delta_{(1)}(\omega_{\rm c})\simeq41.4190M\omega_{\rm c} +\pi.
 \label{constphase2}
\end{align}
The scattered wave in the form of Eq.~\eqref{multi-waves} with $N=3$ is determined by 
Eqs.~\eqref{d-cosine1}--\eqref{d-cosine3}, \eqref{amplitudes}, \eqref{constphase1} and \eqref{constphase2}.

\begin{figure}[htbp]
 \begin{center}
 \includegraphics[width=8cm,clip]{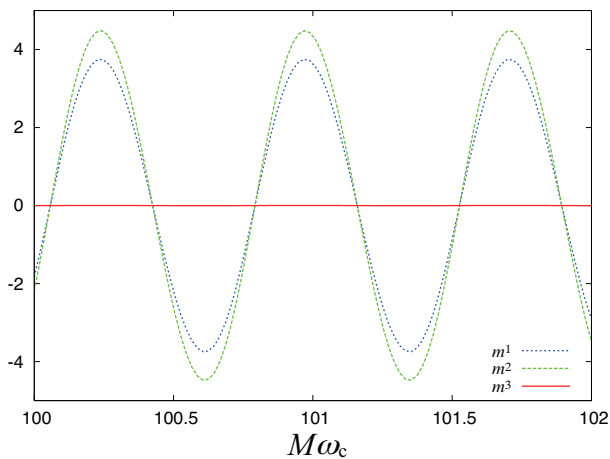}
 \end{center}
 \caption{
 The average values of components of the OAM possessed by one quantum are 
 depicted as a function of $\omegac$ in the case of $a=0.99$. 
 The unit of the OAM is $\hbar$. The third component $m^3$ is much smaller than the other components.  }
 \label{figOAM-Kerr}
\end{figure}

In the previous work by Yang and Casals\cite{Yang:2014kyr}, 
the only $y^3$-component of the OAM was investigated. 
However, we should note that, in general, the other components of the OAM do not vanish. 
We assume 
$$
 \ell_1=\ell_2=\ell_3=100\omegac^{-1},
$$
and depict the numerical results of $m^i$ as a function of $\omegac$ in Fig.\ref{figOAM-Kerr}. 

It is very difficult to see the value of $m^3$ in Fig.~\ref{figOAM-Kerr}. The maximum of $|m^3|$ is 
$9.6\times10^{-3}$. Hence we may say that 
OAM is almost orthogonal to $y^3$-direction which almost agrees 
with the propagation direction of the scalar wave.  
Our results cannot be directly compared with that of Yang and Casals, 
since the present normalization is different from theirs. 
However, very small value of the $y^3$-component of the OAM 
is true in both studies by Yang and Casals and by us. 
Although $|m^3|$ is very small, the norm $\sqrt{\sum_i(m^i)^2}$ is not necessarily so.  

\begin{figure}[htbp]
 \begin{center}
 \includegraphics[width=8cm,clip]{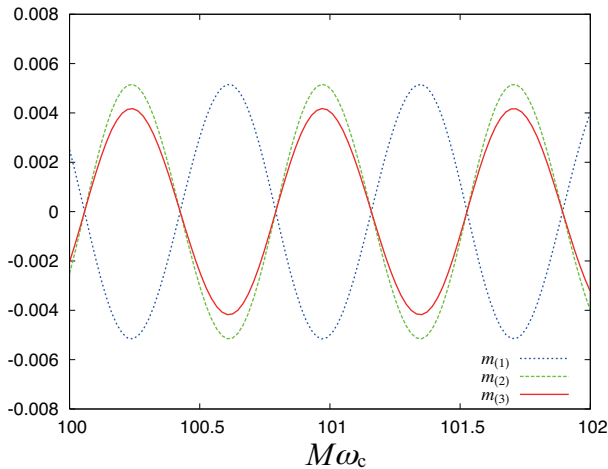}
 \end{center}
 \caption{
 The average values of direction-cosine components of the OAM possessed by one quantum 
 is depicted as a function of $\omegac$ in the case of $a=0.99$. 
 The unit of the OAM is $\hbar$.  }
 \label{figOAM-Kerr-gamma}
\end{figure}

The average values of components of the OAM parallel to the direction cosines of the images have a 
special interest, since the direction cosines of the images are observables. They are defined as
$$
m_{(n)}:=\sum_{i=1}^3\gamma_{(n)}^i m^i. 
$$ 
We depict them as functions of $\omegac$ in Fig.~\ref{figOAM-Kerr-gamma}. The values are too small to 
get values of these components through the measurement of the OAM of the small number of 
quanta, since the OAM of one quantum should take an integer in the unit of $\hbar$. 

\subsection{The case of a non-rotating black hole}

We also investigate the case of a non-rotating black hole, i.e., $a=0$, which is the Schwarzschild spacetime. 
The procedures we should perform are the same as in the case of a rotating black hole. 

We put the emitter and the detector at the following points;
\begin{eqnarray}
 (\theta_{\rm e},\varphi_{\rm e},r_{\rm e})=(\pi/2,\pi,10^4M),~~{\rm and}~~
 (\theta_{\rm d},\varphi_{\rm d},r_{\rm d})\simeq (1.56809,1.98506\times10^{-2},10^4M).
 \label{locations}
\end{eqnarray}
It is worthwhile to notice that the coordinates of the emitter and the detector in this case are the same 
as those in the case of the rotating black hole. 
Then the direction cosines of the brightest three rays are given by
\begin{align}
\gamma^i_{(1)}&=\left(2.71644\times10^{-3},~-1.99209\times10^{-2},~0.999798\right), \cr
\gamma^i_{(2)}&=\left(-1.37635\times10^{-3},~1.00934\times10^{-2},~0.999948\right), \cr
\gamma^i_{(3)}&=\left(-7.02310\times10^{-5},~5.15044\times10^{-4},~1.00020\right). \nonumber
\end{align}
As expected, we can see from the above results that the three direction cosines are not linearly independent 
within the numerical accuracy.  

The ratios of the amplitudes $C_{(n)}$ in Eq.~\eqref{multi-waves} are given by 
\begin{eqnarray}
 \frac{C_{(2)}}{C_{(1)}}= 2.50549\times 10^{-1},~~~~{\rm and}~~~~
 \frac{C_{(3)}}{C_{(1)}}= 2.07062\times 10^{-9}.
 \label{amplitudes-non}
\end{eqnarray}

The numerical results show 
\begin{eqnarray}
 \mathcal{N}_{(1)}=0,~~\mathcal{N}_{(2)}=1,~~{\rm and}~~\mathcal{N}_{(3)}=3,
 \label{caustics-non}
\end{eqnarray}
and hence we have the phase differences as follows; 
\begin{align}
 &\delta_{(2)}(\omega_{\rm c})-\delta_{(1)}(\omega_{\rm c})\simeq
 5.84674~M\omega_{\rm c} +\frac{\pi}{2} 
 \label{constphase1-non}\\
 & \delta_{(3)}(\omega_{\rm c})-\delta_{(1)}(\omega_{\rm c})\simeq 52.0390~M\omega_{\rm c} 
 +\frac{3}{2}\pi.
 \label{constphase2-non}
\end{align}
The scattered wave in the form of Eq.~\eqref{multi-waves} with $N=3$ is determined by 
Eqs.~\eqref{d-cosine1}--\eqref{d-cosine3}, \eqref{amplitudes}, \eqref{constphase1} and \eqref{constphase2}.

\begin{figure}[htbp]
 \begin{center}
 \includegraphics[width=8cm,clip]{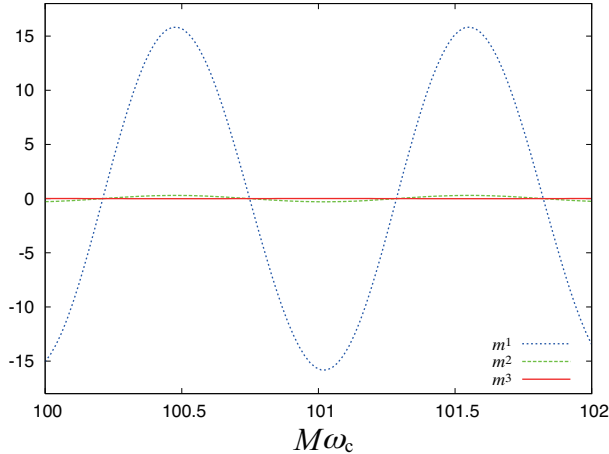}
 \end{center}
 \caption{
 The same as Fig.~\ref{figOAM-Kerr}, but $a=0$. }
 \label{figOAM-Sch}
\end{figure}

As in the case of the rotating black hole, we assume 
$$
 \ell_1=\ell_2=\ell_3=100\omegac^{-1},
$$
and depict the numerical results of $m^i$ as a function of $\omegac$ in Fig.~\ref{figOAM-Sch}. 

Since the Schwarzschild spacetime is spherically symmetric, 
all rays from the emitter to the detector lie on an equatorial plane. 
This means that the direction cosines of all null geodesics are confined in a 2-dimensional subspace 
of the 3-dimensional tangent space orthogonal to $dy^0$; 
As in Sec.~\ref{sec3}-E, we refer the subspace spanned by $\gamma^i_{(m)}$ 
as $S$.  We can see from Eq.~(\ref{Th-1}) 
that the projection of $m^i$ to $S$ necessarily vanishes. 
This fact implies that it is, in principle, possible to know 
through observations of the components of the OAM parallel to 
the direction cosines, $\gamma^i_{(m)}$, whether the black hole is rotating. 
The component of the OAM parallel to the direction cosine of the source of gravity, 
i.e., the $y^3$-direction, also vanishes, since it is an element of $S$. 
This result is consistent with Yang and Casals~\cite{Yang:2014kyr}. 
However, here, we should note that, even in the case of $a=0$, 
the absolute value of $m^i$ is comparable to that in the case of $a=0.99M$.  

\section{Summary and Discussion}\label{sec5}

We considered a situation in which an emitter radiates spherical waves 
of the real massless scalar field 
and revealed a mechanism which generates the orbital angular momentum 
of the scalar wave through gravitational effects. 
As an example of a source of gravity, we considered a black hole. 
We solved the equation of motion of the massless scalar field in the curved spacetime 
by invoking the geometrical optics approximation.
The basic equations of the geometrical optics are the geodesic equations 
with the null condition on the geodesic tangent and the transport equation: 
The former determines the trajectory of the wave, whereas the latter determines the 
amplitude of the wave. Due to the gravitational lensing effects,  
there are infinite number of null geodesics from the emitter to the detector and 
there are infinite number of corresponding ``plane waves" which propagate along 
those null geodesics. The amplitudes of them are very different from each other at the 
detector. We take into account only three null geodesics along which locally plane waves with  
the largest, the second largest and the third largest amplitude at the detector. 

We showed that the superposition of the two locally plane waves at the detector is 
sufficient for the generation of the orbital angular momentum. However, in order to find 
the effect of the frame dragging caused by the rotation of the black hole 
through the detection of the orbital angular momentum, the superposition of the only two 
is not sufficient and the more than two rays are necessary.   
In the case of the non-rotating black hole, the components of the orbital angular momentum 
parallel to the direction cosines of all images do vanish.  
By contrast, if the black hole is rotating, all of them do not vanish due to the frame dragging 
which makes the direction cosines of three images linearly independent.  
Our result may be available 
to the electromagnetic waves. 
Although it seems to be very challenging to observe them, 
our result suggests that the measurement of the orbital angular momentum 
of the electromagnetic waves can be a probe of the spin parameter of the rotating black hole, in principle. 

Finally, we would like to stress that if there is no gravitational lensing effect, or in other words, 
if there is only one image, the orbital angular momentum identically vanishes (see Eq.~(\ref{L-detect})). 
This fact implies that if we detect the non-vanishing orbital angular momentum 
of the electromagnetic wave, 
we may conclude that the multiple images comes from a single emitter or the magnification of 
the apparent luminosity occurs by the gravitational lensing effect. We should note that 
it is not easy task to see whether the multiple images or magnification of the apparent luminosity 
comes from the gravitational lensing effect. The detection of the orbital angular momentum 
may give us a novel criterion for the occurrence of the gravitational lensing.

\section*{Acknowledgments}
KN was supported in part by JSPS KAKENHI Grant Number 25400265. 
YN was supported in part by the JSPS KAKENHI Grant Number 15K05073. 
HI was supported in part by JSPS KAKENHI Grant Number 24540282. 



\end{document}